\documentclass[prl,%
 reprint,
superscriptaddress,
 amsmath,amssymb,
 aps,
]{revtex4-2}

\usepackage{dcolumn}
\usepackage{bm}
\usepackage{amsmath}
\usepackage{braket}
\usepackage{physics}
\usepackage{xcolor}
\usepackage{graphicx}

\begin{document}

\preprint{APS/123-QED}

\title{Scalable Acceleration of Many-Body Quantum Dynamics via Time-Rescaling}

\author{Edson B. de Almeida Filho}
\affiliation{%
Departamento de F\'{i}sica, Universidade Federal da Para\'{i}ba,
58051-900 Jo\~{a}o Pessoa, PB, Brazil
}%

\author{Ângelo F. da Silva França}
\affiliation{%
Departamento de Física, Universidade Estadual da Paraíba,
58429-500 Campina Grande, Paraíba, Brazil
}%

\author{Bertúlio de Lima Bernardo}
\affiliation{%
Departamento de F\'{i}sica, Universidade Federal da Para\'{i}ba,
58051-900 Jo\~{a}o Pessoa, PB, Brazil
}%

\date{\today}

\begin{abstract}

Fast quantum control is essential to overcome decoherence in contemporary quantum platforms, yet achieving this in many-body systems remains a major challenge. We show that the time-rescaling (TR) method enables efficient acceleration of closed many-body quantum dynamics, extending its applicability beyond previously studied regimes. Applying TR to the transverse-field Ising model with a longitudinal field, we demonstrate a significant enhancement of quantum annealing performance, maintaining high ground-state fidelity at evolution times where standard adiabatic dynamics breaks down, with only weak dependence on system size. We further demonstrate high-fidelity preparation of Greenberger–Horne–Zeilinger states in many-body systems, where TR extends the accessible system sizes within fixed evolution times. We additionally show that the Mandelstam-Tamm quantum speed limit does not fundamentally limit the acceleration achievable through TR, as the reduction in evolution time is exactly compensated by increased energy fluctuations. These results establish TR as a scalable and experimentally viable approach to fast quantum control in many-body systems.

\end{abstract}

                    
\maketitle


Recent advances in experimental platforms such as ultracold atoms \cite{aide,gross,ebadi}, trapped ions \cite{friss,zhang}, and superconducting circuits \cite{barends,ma,eick,google,king} have enabled the controlled realization of interacting quantum many-body systems and opened new prospects for quantum computation and simulation \cite{henri,haffner,kja}. These developments place stringent demands on the precise manipulation of quantum states, particularly in regimes where coherence times are limited and system sizes are rapidly increasing. A widely used paradigm for robust quantum control is based on adiabatic evolution, where a system remains in an instantaneous eigenstate of a slowly varying Hamiltonian \cite{messiah,albash}. This approach underlies applications ranging from quantum annealing to state preparation and quantum thermodynamics \cite{campo,odelin}. However, the adiabatic condition generally requires long evolution times, which scale unfavorably with system size and often exceed experimentally accessible coherence windows. As a consequence, purely adiabatic protocols become impractical in many relevant scenarios.

To overcome this limitation, a broad class of techniques known as Shortcuts to Adiabaticity (STA) has been developed to reproduce adiabatic outcomes in finite time \cite{odelin,torrontegui}. Among these, counterdiabatic driving (CD) provides a formally exact solution by supplementing the system Hamiltonian with auxiliary terms that suppress nonadiabatic transitions \cite{demi,demi2,berry}. In its original formulation, the construction of the CD Hamiltonian requires knowledge of the instantaneous eigenstates and eigenvalues of the system. Modern formulations recast this problem in terms of the adiabatic gauge potential, which is generally unknown and highly nonlocal in interacting many-body systems \cite{pandey}. These challenges have motivated the development of variational approaches \cite{sels}, in which the gauge potential is approximated within a restricted operator manifold. Such methods substantially extend the applicability of CD without requiring explicit spectral information \cite{sels,kolo,claeys,cepaite}. Nevertheless, they generally provide only approximate suppression of nonadiabatic transitions, rely on detailed knowledge of the underlying Hamiltonian, and require additional control terms whose implementation may become increasingly challenging as the system size grows \cite{taka,bhatt,chanda,hega,wurtz}. As a result, the realization of CD protocols in large interacting systems remains a significant challenge \cite{ohga}.

An alternative route is provided by the time-rescaling (TR) method, introduced in Ref.~\cite{bernardo}, which accelerates quantum dynamics through a reparameterization of the time variable. This approach has been successfully applied in a variety of settings, including few-level systems, continuous-variable dynamics, relativistic regimes, and open quantum systems \cite{andrade,lukas,roy,bernardo2}. In contrast to conventional STA techniques, TR preserves the structure of the original Hamiltonian and does not require auxiliary control fields or knowledge of instantaneous eigenstates. Despite these advantages, the applicability of TR to closed many-body quantum systems, where scalability constitutes the central challenge, remains an open question.

In this work, we demonstrate that TR provides an effective and scalable strategy for accelerating many-body quantum dynamics. Focusing on the transverse-field Ising model with a longitudinal field, we show that TR significantly enhances the performance of quantum annealing protocols, maintaining high ground-state fidelity at evolution times where standard adiabatic dynamics breaks down. Importantly, this advantage persists with increasing system size, indicating robustness beyond the few-body regime. We further apply the method to the preparation of Greenberger–Horne–Zeilinger (GHZ) states, achieving high-fidelity generation using only temporal reparameterization of the driving protocol. In addition, we show that the acceleration induced by TR remains fully compatible with the Mandelstam-Tamm quantum speed limit (QSL), since the reduction in evolution time is exactly compensated by increased energy fluctuations. Our results thus open new possibilities for the implementation of fast quantum protocols in current and near-term experimental platforms.

{\it Time-rescaling framework.—}
To address the acceleration of many-body quantum dynamics, we employ the TR method~\cite{bernardo}. For completeness, we briefly recall its essential ingredients, focusing on the aspects relevant to the present work. We consider a quantum system governed by a time-dependent Hamiltonian $H(t)$, which generates the reference dynamics over a total duration $t_f$. The TR method enables the acceleration of this evolution through a reparameterization of time. Specifically, by introducing a monotonic mapping $t \rightarrow f(t)$, the dynamics can be equivalently described by an effective Hamiltonian
\begin{equation}
\mathcal{H}(t) = H[f(t)]\,\dot{f}(t),
\end{equation}
which preserves the structure of the original Hamiltonian while modifying the rate at which the protocol is traversed. As a consequence, the time-rescaled dynamics reaches the same final state as the reference evolution in a shorter time.

To ensure that the system experiences the same physical conditions as in the reference protocol at the boundaries of the accelerated evolution, the TR function $f(t)$ must preserve the initial and final Hamiltonians. A convenient choice is given by \cite{bernardo}
\begin{equation}
f(t) = a t - \frac{(a-1)}{2 \pi a} t_{f} \sin \left( \frac{2 \pi a}{t_{f}} t \right),
\end{equation}
where $a>1$ is the time-contraction (acceleration) parameter. This function generates a shortened evolution with duration $\tau = t_f/a$ while preserving smooth boundary conditions. A central feature of the TR protocol is that it preserves the dynamical route of the system in Hilbert space \cite{lukas,bernardo2}. Indeed, the accelerated evolution satisfies $\ket{\tilde{\psi}(t)} = \ket{\psi[f(t)]}$, where $\ket{\psi(t)}$ denotes the reference evolution (see Suplemental Material \cite{SM}). Consequently, the system follows the same trajectory in Hilbert space as in the original dynamics, but traversed at a different rate. The probability amplitudes in the instantaneous basis therefore remain identical up to the time reparameterization. In particular, if the reference dynamics is transitionless, the corresponding TR protocol is also transitionless, while reaching the same final state in a shorter time. In the following, we apply the TR method to accelerate quantum annealing protocols in many-body quantum systems.

{\it Quantum annealing acceleration.—}
In its general formulation, quantum annealing is described by a time-dependent Hamiltonian \cite{das,tanaka,albash,hauke,rajak}
\begin{equation}
H(t) = A(t) H_D + B(t) H_P,
\end{equation}
where $H_D$ is a driver Hamiltonian, whose ground state is easy to prepare, and $H_P$ is a problem Hamiltonian, whose ground state encodes the target solution \cite{lucas}. In standard quantum annealing protocols, the schedules $A(t)$ and $B(t)$ are generally chosen such that $A(0)\gg B(0)$ and $A(t_f)\ll B(t_f)$, ensuring a continuous interpolation between the driver and problem Hamiltonians. In the adiabatic limit, this evolution prepares the target ground state with high fidelity. However, for finite evolution times, nonadiabatic transitions become significant, particularly in regions where the energy gap is minimal \cite{das}. This effect is exacerbated in many-body systems, leading to a rapid degradation of the annealing performance \cite{wurtz,farhi}. To overcome this limitation, we apply the TR protocol introduced above, replacing $H(t)$ by the effective Hamiltonian $\mathcal{H}(t)=H[f(t)]\dot{f}(t)$, and investigate its ability to accelerate the annealing dynamics.

{\it Ising chain with transverse and longitudinal fields.—}
We now consider a specific realization of the quantum annealing protocol introduced above, based on the Hamiltonian of the one-dimensional Ising model in transverse and longitudinal fields,
\begin{equation}
H(t) = -J \sum_{j=1}^{N-1} \sigma_j^z \sigma_{j+1}^z + h_z \sum_{j=1}^{N} \sigma_j^z + \lambda(t) h_x \sum_{j=1}^{N} \sigma_j^x,
\end{equation}
where $\sigma_j^{x,z}$ are Pauli operators, $h_z$ is the strength of the longitudinal field, $h_x$ is the final strength of the transverse field, and $N$ is the number of spins. This Hamiltonian can be represented in the general form $H(t)=A(t)H_D+B(t)H_P$, with $H_D= h_x \sum_j \sigma_j^x$, $H_P=-J \sum_{j} \sigma_j^z \sigma_{j+1}^z + h_z \sum_{j} \sigma_j^z$, $A(t)=\lambda(t)$, and $B(t)=1$. The function $\lambda(t)$ defines the annealing schedule, satisfying $\lambda(0)=0$ and $\lambda(t_f)=1$. In this work, we adopt the smooth protocol
\begin{equation}
\label{schedule}
\lambda(t) = \sin^{2} \left[ \frac{\pi}{2} \sin^{2} \left( \frac{\pi t}{2 t_f} \right) \right],
\end{equation}
which ensures vanishing derivatives at the boundaries.

\begin{figure*}
    \centering
\includegraphics[width=0.75\textwidth]{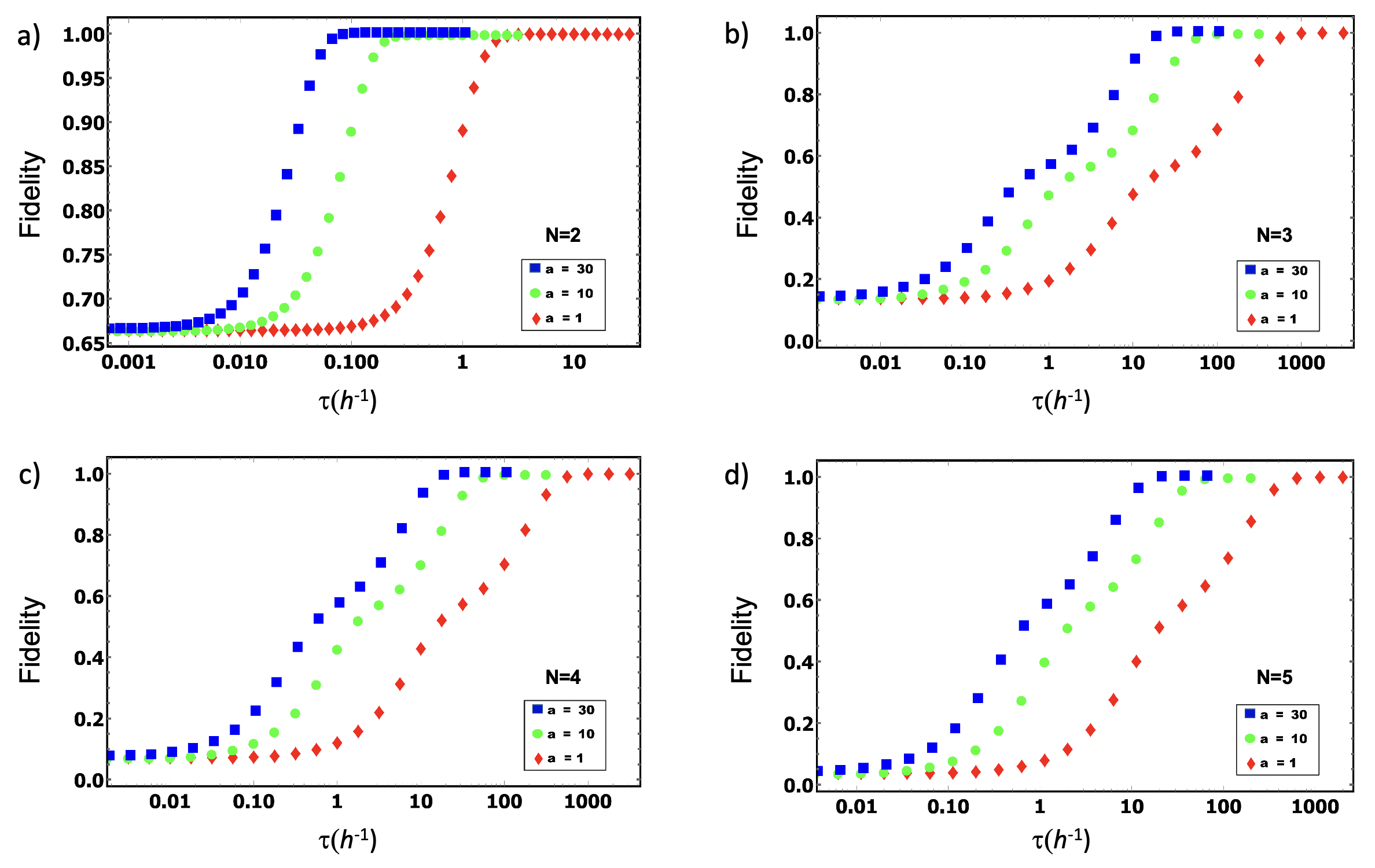}
    \centering
    \caption{
Final-state fidelity $\mathcal{F}$ as a function of the evolution time $\tau$ (logarithmic scale) for Ising chains with transverse and longitudinal fields containing (a) $N=2$, (b) $N=3$, (c) $N=4$, and (d) $N=5$ spins. The reference protocol ($a=1$, red diamonds) is compared with time-rescaled dynamics for $a=10$ (green circles) and $a=30$ (blue squares). Increasing the time-contraction parameter $a$ shifts the high-fidelity regime toward shorter evolution times, enabling accurate state preparation in parameter regions where the standard annealing protocol fails. Parameters: for $N=2$, $J=h$, $h_z=-h$, and $h_x=2h$; for $N=3,4,5$, $J=h$, $h_z=0.02h$, and $h_x=10h$.}
\label{Fig1}
\end{figure*}

The system is initialized in the ground state of the Ising Hamiltonian with a longitudinal field, corresponding to a fully polarized configuration ($\vert \uparrow\uparrow\cdots\uparrow \rangle$), and is driven toward the ground state of the full interacting Hamiltonian including the transverse field, which is given by a superposition of basis states. The parameters specified below define the reference (unaccelerated) annealing protocols, which are subsequently accelerated via the TR transformation. For the elementary two-spin system, we adopt $J=h$, $h_z=-h$, and $h_x=2 h$, where $h>0$ define the interaction and field energy scales. For larger systems ($N=3,4,5$), we consider $J=h$, $h_z=0.02h$, and $h_x=10h$. The implementation of the TR transformation, defined by the effective Hamiltonian $\mathcal{H}(t)=H[f(t)]\dot{f}(t)$, requires a modified control of the transverse field, governed by the rescaled protocol $\tilde{\lambda}(t)=\lambda[f(t)]\dot{f}(t)$.

The performance of the protocol is quantified by the final-state fidelity $\mathcal{F}$ between the evolved state and the target ground state. Figure~\ref{Fig1} shows $\mathcal{F}$ as a function of the total evolution time $\tau$ for different values of the time-contraction parameter $a$. While the reference evolution ($a=1$) requires long times to reach high fidelity, the TR protocol achieves $\mathcal{F}\approx1$ at significantly shorter times. As shown in Fig.~\ref{Fig1}, increasing $a$ produces an approximately horizontal shift of the fidelity curves toward shorter times, indicating that the overall structure of the dynamics is preserved while its characteristic timescale is systematically reduced. Remarkably, the accelerated protocols ($a>1$) consistently achieve fidelities above $99\%$ across all system sizes, even in regimes where adiabatic evolution breaks down.

\begin{figure}[hbt!]
\begin{center}
\includegraphics[width=0.45\textwidth]{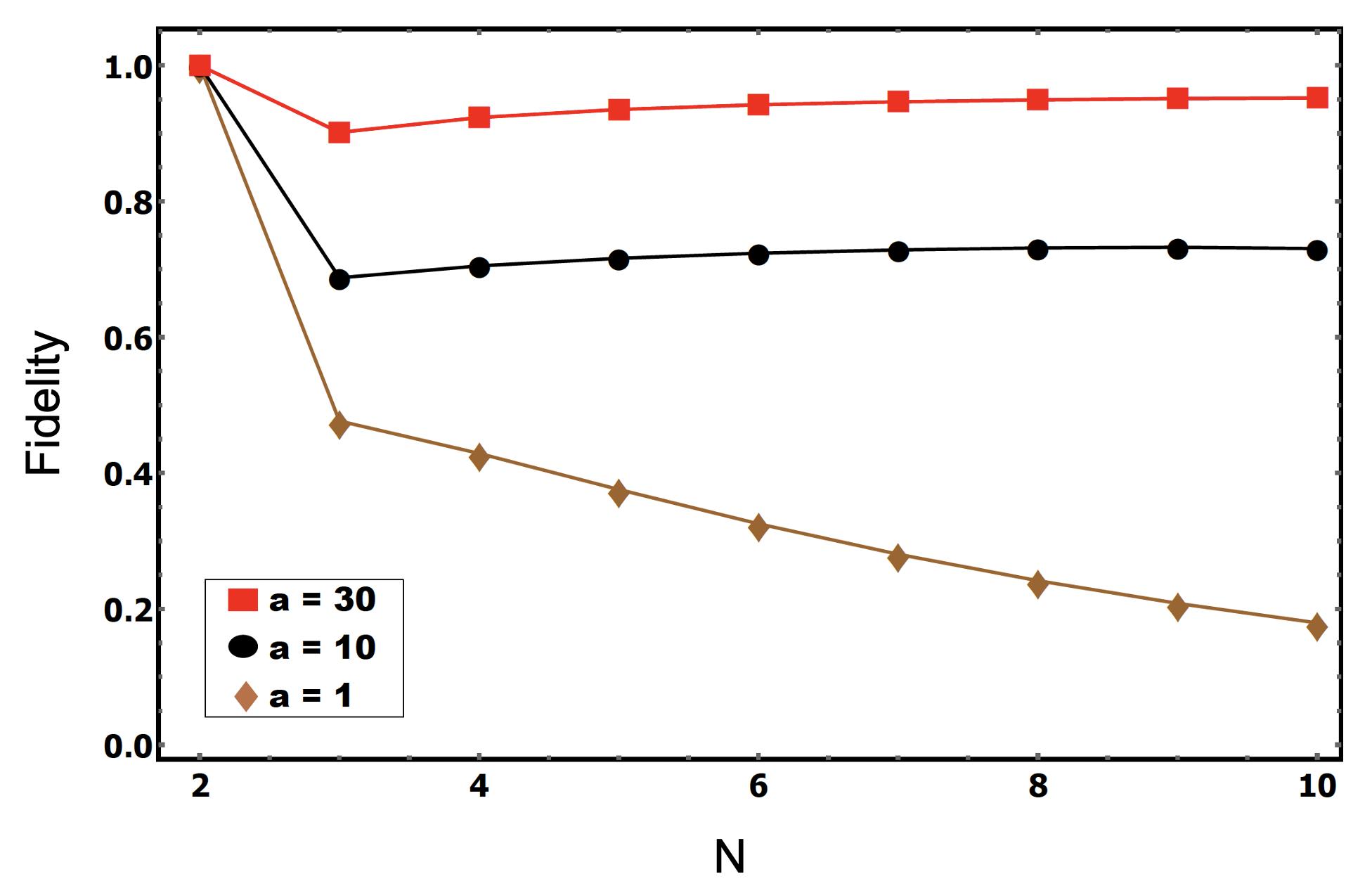}
\caption{
Final-state fidelity $\mathcal{F}$ as a function of system size $N$ at fixed evolution time $\tau = 10$, for different values of the time-contraction parameter $a$. The reference protocol ($a=1$, red diamonds) exhibits a pronounced decay of fidelity as the system size increases, reflecting the progressive breakdown of adiabaticity in the many-body regime. In contrast, the time-rescaled dynamics ($a=10$, green circles; $a=30$, blue squares) maintains substantially higher fidelity with only weak dependence on $N$. In particular, for $a=30$, the fidelity remains close to unity even for $N=10$, demonstrating the robustness of the TR protocol against increasing system size. Parameters used in the simulations are $J=1$, $h_z=0.02$ and $h_x=10$ for all values of $N$.
}
\label{Fig2}
\end{center}
\end{figure}

We next analyze the behavior as a function of system size. Figure~\ref{Fig2} shows the fidelity at fixed evolution time for increasing $N$. The reference protocol exhibits a rapid degradation of performance with system size, reflecting the enhanced role of nonadiabatic excitations in larger systems. In contrast, the time-rescaled dynamics maintains high fidelity across the range of system sizes considered, with only a weak dependence on $N$. Notably, for $a=30$, the fidelity remains nearly constant with increasing system size over the time interval considered, indicating a strong suppression of the size-dependent breakdown of adiabaticity. These results indicate that TR provides a robust route to fast quantum annealing in many-body systems.

In Ref.~\cite{cepaite}, the acceleration of the annealing dynamics described above was investigated using local counterdiabatic driving (LCD) and counterdiabatic optimized local driving (COLD), the latter combining LCD with quantum optimal control techniques. In order to achieve high-fidelity accelerated dynamics, the LCD approach required the introduction of additional nonlocal long-range interaction terms, whereas the COLD protocol relied on extra local Hamiltonian contributions with intricate time-dependent modulations. In contrast, the TR protocol generates accelerated dynamics solely through the modulation of control parameters already present in the original Hamiltonian, while involving a substantially simpler temporal structure.

{\it GHZ state preparation.—}
We consider the preparation of a Greenberger–Horne–Zeilinger (GHZ) state \cite{GHZ}, $\frac{1}{\sqrt{2}}\left( |\uparrow\uparrow\cdots\uparrow\rangle + |\downarrow\downarrow\cdots\downarrow\rangle \right)$,
in a one-dimensional Ising chain. To this end, we employ a quantum annealing protocol based on the Hamiltonian
\begin{equation}
\label{GHZsp}
H(t) = (1 - \lambda(t))\, h \sum_{i=1}^{N} \sigma_i^x + \lambda(t)\, J \sum_{i=1}^{N} \sigma_i^z \sigma_{i+1}^z,
\end{equation}
with ferromagnetic coupling $J<0$ (here fixed to $J=-1$) and periodic boundary conditions, $\sigma_{N+1}=\sigma_1$. The annealing schedule $\lambda(t)$ is defined in Eq.~(\ref{schedule}) and interpolates between the transverse field and interaction terms. At the initial time, $\lambda(0)=0$, the system is prepared in the ground state of the transverse field Hamiltonian, $\ket{+\cdots+}$, where $\ket{+}=(\ket{\uparrow}+\ket{\downarrow})/\sqrt{2}$, corresponding to a product state polarized along the $x$ direction ($h>0$). As $\lambda(t)$ increases, the transverse field is suppressed while the Ising interaction is turned on. At the final time, $\lambda(t_f)=1$, the Hamiltonian reduces to the ferromagnetic Ising model, whose ground-state manifold is twofold degenerate. The GHZ state is obtained as the symmetric superposition of these degenerate ferromagnetic ground states.

\begin{figure}[hbt!]
\begin{center}
\includegraphics[width=0.45\textwidth]{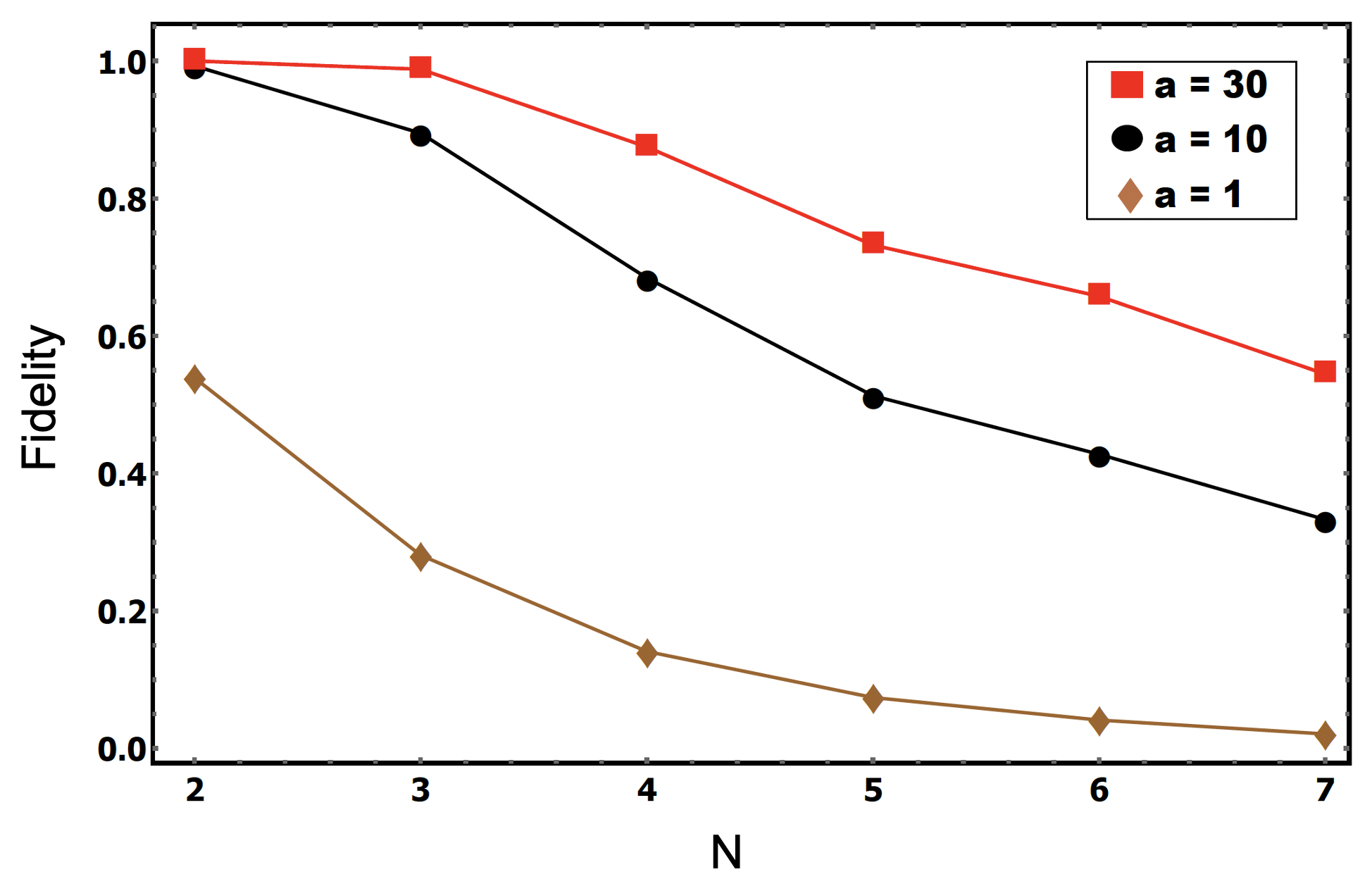}
\caption{
Final-state fidelity $\mathcal{F}$ for GHZ state preparation as a function of system size $N$ at fixed evolution time $\tau = 1$. The protocol is implemented using the Hamiltonian in Eq.~(\ref{GHZsp}) with $J=-1$, $h=-1/2$, and periodic boundary conditions. The reference protocol ($a=1$, brown diamonds) exhibits a rapid decay of fidelity with increasing $N$, approaching zero for $N \gtrsim 5$. In contrast, the time-rescaled dynamics ($a=10$, black circles; $a=20$, red squares) significantly enhances the fidelity and extends the range of system sizes over which useful GHZ states can be generated. The achieved performance is comparable to that of Ref.~\cite{sun2022}, despite requiring only temporal reparameterization.
}
\label{Fig3}
\end{center}
\end{figure}

Figure~\ref{Fig3} shows the final state fidelity $\mathcal{F}$ as a function of system size at fixed evolution time. The reference protocol ($a=1$) exhibits a rapid breakdown of performance, with the fidelity dropping to near zero for moderate system sizes, indicating the failure of adiabatic state preparation for highly entangled states in short times. In contrast, the time-rescaled dynamics significantly enhances $\mathcal{F}$ and systematically extends the range of system sizes over which useful GHZ states can be generated. While moderate acceleration ($a=10$) partially mitigates the decay, larger values of the time-contraction parameter ($a=20$) sustain substantially higher fidelities across the system sizes considered. This behavior demonstrates that TR delays the onset of the size-dependent breakdown and enables the preparation of entangled states in regimes that are inaccessible to the standard protocol.

A related problem was investigated in Ref.~\cite{sun2022} using variational quantum circuits to optimize counterdiabatic (CD) coefficients. In that approach, the accelerated dynamics required additional time-dependent two-body interactions already at lowest order, while higher-order corrections generated increasingly complex many-body terms that are difficult to implement experimentally. Despite this substantially higher control complexity, the achieved fidelities are comparable to those obtained here. This comparison underscores a central advantage of the TR protocol: the acceleration is realized exclusively through temporal reparameterization, without auxiliary control fields or engineered many-body interactions. These results identify TR as an efficient and experimentally accessible approach for the fast generation of highly entangled many-body states.

{\it Quantum speed limit.—}
The acceleration achievable through TR is accompanied by increased energy fluctuations, as generally occurs in STA protocols \cite{odelin,poggi}. For the TR Hamiltonian, $\mathcal{H}(t)=H[f(t)]\dot f(t)$, the instantaneous energy uncertainty satisfies $\Delta\mathcal{E}(t)=\dot f(t)\Delta E_{\rm ref}[f(t)]$, where $\Delta E_{\rm ref}$ is the energy uncertainty associated with the reference evolution. As a consequence, the time-averaged uncertainty scales as $\overline{\Delta\mathcal{E}}=a\,\overline{\Delta E}_{\rm ref}$, while the protocol duration is reduced to $\tau=t_f/a$ (see Supplemental Material \cite{SM}). Consequently, the increase in energetic uncertainty exactly compensates the time compression in the Mandelstam--Tamm QSL bound \cite{MT,DL,deffner}. Therefore, the QSL does not impose an intrinsic upper bound on the time-contraction parameter $a$. Instead, similarly to CD and related STA approaches \cite{bukov}, increasingly fast protocols require proportionally larger physical resources, including stronger control fields and enhanced energy fluctuations. The achievable acceleration via TR is thus ultimately limited not by the QSL itself, but by experimental constraints such as finite field strengths, limited modulation bandwidth, decoherence, and noise.

{\it Discussion.—} We have shown that TR provides an efficient and experimentally accessible route to accelerate quantum dynamics in many-body systems. By reparameterizing the temporal evolution, the method achieves high-fidelity state preparation in regimes where standard adiabatic dynamics fails, without requiring auxiliary control Hamiltonians or spectral information. Our results demonstrate three key features. First, TR enables a substantial acceleration of quantum annealing, shifting the dynamics to significantly shorter times while preserving the structure of the evolution. Second, the protocol mitigates the degradation of performance with increasing system size, maintaining high fidelity in regimes where adiabatic evolution breaks down. Third, the TR protocol applies to distinct classes of many-body quantum tasks, including ground-state preparation and the generation of highly entangled states, exemplified by the GHZ protocol. In both cases, TR significantly enlarges the range of system sizes accessible within fixed evolution times.

A central advantage of the TR approach lies in its simplicity: acceleration is achieved solely through a modification of the time parameter, without introducing additional interactions or requiring knowledge of instantaneous eigenstates, in contrast to CD. This feature makes the method particularly attractive for implementation in complex many-body systems and near-term quantum hardware. We further showed that the Mandelstam-Tamm QSL does not impose an intrinsic upper bound on the time-contraction parameter $a$, since the reduction in evolution time is exactly compensated by increased energy fluctuations. Thus, similarly to CD-based STA protocols, the achievable acceleration via TR is ultimately limited by experimental resources rather than by fundamental quantum dynamical bounds. Accordingly, the accelerated many-body protocols presented here may, in principle, be implemented on even shorter timescales, provided sufficiently strong and precise control resources are available.

{\it Acknowledgement.—} The authors acknowledge the support of Coordenação de Aperfeiçoamento de Pessoal de Nível Superior (CAPES, Finance Code 001) and Conselho Nacional de Desenvolvimento Científico e Tecnológico (CNPq). B.L.B. acknowledges support from CNPq (Grant No. 307876/2022-5).

\clearpage
\onecolumngrid

\begin{center}
\textbf{\large Supplemental Material for ``Scalable Acceleration of Many-Body Quantum Dynamics via Time-Rescaling''}
\end{center}

\begin{quotation}
In this Supplemental Material, we show that the time-rescaling (TR) protocol preserves the Hilbert-space trajectory of the reference evolution while compressing its duration. We also demonstrate that TR is fully consistent with the Mandelstam–Tamm quantum speed limit, as the reduction in evolution time is exactly compensated by a proportional increase in the time-averaged energy uncertainty.
\end{quotation}

\section{Time-Rescaled Dynamics}

We consider a reference quantum evolution generated by a time-dependent Hamiltonian $H(t)$ over a total duration $t_f$. The corresponding state $\ket{\psi(t)}$ satisfies the Schr\"odinger equation
\begin{equation}
i\hbar \frac{\partial}{\partial t} \ket{\psi(t)}
=
H(t)\ket{\psi(t)},
\label{S1}
\end{equation}
with $0\le t \le t_f$. The time-rescaling (TR) protocol is constructed through a monotonic reparameterization of time, $t\rightarrow f(t)$, such that the accelerated dynamics follows the same trajectory in Hilbert space as the reference evolution, but traversed at a different rate \cite{bernardosm}. The state associated with the accelerated process is defined as
\begin{equation}
\ket{\tilde{\psi}(t)}
=
\ket{\psi[f(t)]},
\label{S3}
\end{equation}
where the rescaling function satisfies the boundary conditions $f(0)=0$ and
$f(\tau)=t_f$, where
$\tau = t_f / a$
denotes the duration of the accelerated protocol, with $a>1$ defining the time-contraction parameter. Substituting Eq.~(\ref{S3}) into the Schr\"odinger equation, one finds that the accelerated evolution is generated by the Hamiltonian
\begin{equation}
\mathcal{H}(t)
=
H[f(t)]\dot f(t),
\label{S6}
\end{equation}
where $\dot f(t)\equiv df/dt$.

To demonstrate this result, we differentiate Eq.~(\ref{S3}) with respect to time. Using the chain rule, one obtains
\begin{equation}
\frac{d}{dt}\ket{\tilde{\psi}(t)} 
= \frac{d}{dt}|\psi[f(t)]\rangle =
\dot f(t)
\left.
\frac{d}{ds}\ket{\psi(s)}
\right|_{s=f(t)} .
\label{S6a}
\end{equation}
Multiplying by $i\hbar$ and using the reference Schr\"odinger equation,
\begin{equation}
i\hbar
\frac{d}{ds}\ket{\psi(s)}
=
H(s)\ket{\psi(s)},
\label{S6b}
\end{equation}
evaluated at $s=f(t)$, yields
\begin{equation}
i\hbar
\frac{d}{dt}\ket{\tilde{\psi}(t)}
=
\dot f(t)\,
H[f(t)]
\ket{\psi[f(t)]}.
\label{S6c}
\end{equation}
Finally, using Eq.~(\ref{S3}), we get
\begin{equation}
i\hbar
\frac{d}{dt}\ket{\tilde{\psi}(t)}
=
H[f(t)]\dot f(t)
\ket{\tilde{\psi}(t)},
\label{S6d}
\end{equation}
which proves the result of Eq.~(\ref{S6}).

In the present work, we employ the TR function
\begin{equation}
f(t)
=
a t
-
\frac{(a-1)}{2\pi a}
t_f
\sin\left(
\frac{2\pi a}{t_f}t
\right),
\label{S7}
\end{equation}
whose derivative is
\begin{equation}
\dot f(t)
=
a-(a-1)
\cos\left(
\frac{2\pi a}{t_f}t
\right).
\label{S8}
\end{equation}
This construction satisfies the smooth boundary conditions
\begin{equation}
\dot f(0)=\dot f(\tau)=1,
\label{S9}
\end{equation}
while ensuring
\begin{equation}
\dot f(t)\ge 1
\qquad
(a>1).
\label{S10}
\end{equation}
Accordingly, the time-rescaled dynamics evolves locally faster than the reference process throughout the entire protocol.

\section{Energy Uncertainty Under Time-Rescaling}

We now analyze how the energy uncertainty of the evolution changes under TR. The instantaneous energy uncertainty associated with the accelerated Hamiltonian $\mathcal{H}(t)$ is defined as
\begin{equation}
\Delta \mathcal{E}(t)^2
=
\expval{\mathcal{H}^2(t)}
-
\expval{\mathcal{H}(t)}^2.
\label{S11}
\end{equation}
Using Eq.~(\ref{S6}), we obtain
\begin{equation}
\mathcal{H}^2(t)
=
\dot f(t)^2
H^2[f(t)],
\label{S12}
\end{equation}
which immediately yields
\begin{equation}
\Delta \mathcal{E}(t)
=
|\dot f(t)|
\Delta E_{\mathrm{ref}}[f(t)],
\label{S13}
\end{equation}
where
\begin{equation}
\Delta E_{\mathrm{ref}}(s)
=
\sqrt{
\expval{H^2(s)}
-
\expval{H(s)}^2
}
\label{S14}
\end{equation}
is the instantaneous energy uncertainty of the reference dynamics.

Since the TR function employed here satisfies $\dot f(t)\ge0$, Eq.~(\ref{S13}) simplifies to
\begin{equation}
\Delta \mathcal{E}(t)
=
\dot f(t)\,
\Delta E_{\mathrm{ref}}[f(t)].
\label{S15}
\end{equation}
Equation~(\ref{S15}) shows that the instantaneous energy fluctuations are amplified by the local time-contraction factor $\dot f(t)$. Therefore, the acceleration induced by the TR protocol is necessarily accompanied by an increase in the energy uncertainty of the dynamics.

\section{Time-Averaged Energy Uncertainty}

To establish the connection with the quantum speed limit (QSL), we consider the time-averaged energy uncertainty associated with the accelerated process,
\begin{equation}
\overline{\Delta \mathcal{E}}
=
\frac{1}{\tau}
\int_0^\tau
\Delta \mathcal{E}(t)\,dt.
\label{S16}
\end{equation}
Substituting Eq.~(\ref{S15}) into Eq.~(\ref{S16}) gives
\begin{equation}
\overline{\Delta \mathcal{E}}
=
\frac{1}{\tau}
\int_0^\tau
\dot f(t)
\Delta E_{\mathrm{ref}}[f(t)]
\,dt.
\label{S17}
\end{equation}
We now perform the change of variables
\begin{equation}
s=f(t),
\qquad
ds=\dot f(t)\,dt.
\label{S18}
\end{equation}
Using the boundary conditions $f(0)=0$ and
$f(\tau)=t_f$, we obtain
\begin{equation}
\overline{\Delta \mathcal{E}}
=
\frac{1}{\tau}
\int_0^{t_f}
\Delta E_{\mathrm{ref}}(s)\,ds.
\label{S19}
\end{equation}
Defining the time-averaged energy uncertainty of the reference evolution as
\begin{equation}
\overline{\Delta E}_{\mathrm{ref}}
=
\frac{1}{t_f}
\int_0^{t_f}
\Delta E_{\mathrm{ref}}(s)\,ds,
\label{S20}
\end{equation}
and using that $\tau=t_f/a$, we finally obtain
\begin{equation}
\overline{\Delta \mathcal{E}}
=
a\,
\overline{\Delta E}_{\mathrm{ref}}
.
\label{S21}
\end{equation}
Therefore, the average energy uncertainty grows linearly with the time-contraction parameter $a$. This relation provides a direct quantitative connection between acceleration and energy uncertainty in time-rescaled protocols.

\section{Quantum Speed Limit}

We now analyze the implications of Eq.~(\ref{S21}) for the quantum speed limit (QSL). For unitary dynamics generated by a time-dependent Hamiltonian, the Mandelstam-Tamm bound reads \cite{DL,deffner}
\begin{equation}
\tau
\ge
\frac{
\hbar\,\mathcal{L}
}{
\overline{\Delta \mathcal{E}}
},
\label{S22}
\end{equation}
where
\begin{equation}
\mathcal{L}
=
\arccos\left(
|\braket{\psi_i}{\psi_f}|
\right)
\label{S23}
\end{equation}
is the Bures angle between the initial and final states. Substituting Eq.~(\ref{S21}) into Eq.~(\ref{S22}), we obtain
\begin{equation}
\tau
\ge
\frac{
\hbar\,\mathcal{L}
}{
a\,\overline{\Delta E}_{\mathrm{ref}}
}.
\label{S24}
\end{equation}
Using $\tau=t_f/a$, this expression becomes
\begin{equation}
\frac{t_f}{a}
\ge
\frac{
\hbar\,\mathcal{L}
}{
a\,\overline{\Delta E}_{\mathrm{ref}}
},
\label{S25}
\end{equation}
which provides 
\begin{equation}
t_f
\ge
\frac{
\hbar\,\mathcal{L}
}{
\overline{\Delta E}_{\mathrm{ref}}
}.
\label{S26}
\end{equation}
This is precisely the QSL associated with the original reference evolution.

This analysis demonstrates that the time-rescaled protocol preserves the QSL in a nontrivial and exact manner. Although the total evolution time is reduced by a factor $a$, the time-averaged energy uncertainty increases by the same factor. Hence, the product $\tau\,\overline{\Delta \mathcal{E}}$ remains invariant under TR. This means that the acceleration of the dynamics is achieved through a precise trade-off between evolution time and energy fluctuations. In this sense, the TR transformation realizes a controlled redistribution of the energetic resources required to implement the evolution while preserving the same fundamental dynamical bound as the reference process.

\end{document}